\documentclass{aa}  
\usepackage{graphicx}
\usepackage[breaklinks=true]{hyperref}
\hypersetup{colorlinks=true,urlcolor=blue,citecolor=blue,pdfborder= 0 0 0}
\bibpunct{(}{)}{;}{a}{}{,} 
\usepackage{txfonts}

\begin{document} 

   \title{Multifrequency studies of galaxies and groups: I. Environmental effect on galaxy stellar mass and morphology.\thanks{The multifrequency catalog is available in electronic form at the CDS via anonymous ftp to cdsarc.u-strasbg.fr (130.79.128.5)
or via \url{http://cdsweb.u-strasbg.fr/cgi-bin/qcat?J/A+A/}}}
   \subtitle{}
   \author{A. Poudel
          \inst{1,\star}
          \and
          P. Heinämäki\inst{1}\
          \and
          P. Nurmi\inst{1,2}
      \and
          P. Teerikorpi\inst{1}
       \and             
              E. Tempel\inst{3}
           \and                 
              H. Lietzen\inst{4,5}
       \and             
              M. Einasto\inst{3}                 
          }
           
   \institute{\inst{1} Tuorla Observatory, University of Turku,
              Väisäläntie 20, Piikkiö, Turku, Finland\\
              \inst{2} Finnish Centre for Astronomy with ESO (FINCA), University of Turku,
              Väisäläntie 20, Piikkiö, Turku, Finland\\
              \inst{3} Tartu Observatory, Observatooriumi 1, 61602 T\~oravere, Estonia\\
              \inst{4}Instituto de Astrof\' isica de Canarias, E-38205 La Laguna, Tenerife, Spain\\
              \inst{5}Universidad de La Laguna, Dept. Astrof\' isica, E-38206 La Laguna, Tenerife, Spain\\
              $^{\star}$ \email{anuppou@utu.fi}
              }
   \date{Received ; accepted }
 
  \abstract
   {To understand the role of the environment in galaxy formation, evolution, and present-day properties, it is essential to study the
multifrequency behavior of different galaxy populations under various environmental conditions.
}
   {We study the stellar mass functions of different galaxy populations in groups as a function of their large-scale environments using multifrequency observations.
}
   {We cross-matched the SDSS DR10 group catalog with GAMA Data Release 2 and Wide-field Survey Explorer (WISE) data to construct a catalog of 1651 groups and 11436 galaxies containing photometric information in 15 different wavebands ranging
from ultraviolet (0.152~$\mu$m) to mid-infrared (22~$\mu$m). We performed the spectral energy distribution (SED) fitting of galaxies using the MAGPHYS code and estimate the rest-frame luminosities and stellar masses. We used the $1/V_\mathrm{max}$ method to estimate the galaxy stellar mass and luminosity functions, and the luminosity density field of galaxies to define the large-scale environment of galaxies.}
   {The stellar mass functions of both central and satellite galaxies in groups are different in low- and high-density, large-scale environments. Satellite galaxies in high-density environments have a steeper low-mass end slope compared to low-density environments, independent of the galaxy morphology. Central galaxies in low-density environments have a steeper low-mass end slope, but the difference disappears for fixed galaxy morphology. The characteristic stellar mass of satellite galaxies is higher in high-density environments and the difference exists only for galaxies with elliptical morphologies.}
   {Galaxy formation in groups is more efficient in high-density, large-scale environments. Groups in high-density environments have higher abundances of satellite galaxies, irrespective of the satellite galaxy morphology. The elliptical satellite galaxies are generally more massive in high-density environments. The stellar masses of spiral satellite galaxies show no dependence on the large-scale environment.}


   \maketitle
%
\section{Introduction}
Galaxies and their dark matter halos have grown hierarchically, forming larger systems, such as groups, clusters, filaments, and superclusters of galaxies, separated by enormous voids. In observations, this large-scale distribution of dark and baryonic matter manifests itself as a complex web of galaxies. This cosmic web includes a wide range of cosmic scales from the mega-parsec (Mpc) galaxy group scale to the supercluster and filament scales, several tens to hundreds of Mpc \citep{1978IAUS...79..241J,1978ApJ...222..784G,1979ApJ...230..648C,1982Natur.300..407Z,2014MNRAS.438.3465T}. As a result, galaxies reside in a variety of environments. A natural question is whether the interplay of various physical processes during the hierarchical growth of structures has also affected the properties of galaxies and which scale of environment plays a dominant role in galaxy evolution.

The mass of a galaxy is likely the main driver of its evolution \citep{1996A&A...312..397G, 2001AJ....121..753B, 2002ApJ...576..135G}. Until recently, theoretical models of the halo-galaxy relationship have usually assumed that galaxy populations in dark matter halos depend solely on the halo mass and do not depend on its large-scale environment and assembly history \citep[][and references therein]{2010MNRAS.409..936P}. Recent results, however, suggest that the dependence on halo mass alone cannot explain all galaxy properties. In theoretical studies, older dark matter halos are found to cluster more strongly than more recently formed halos of the same mass \citep{2005MNRAS.363L..66G}. This effect, which is known as assembly bias, suggests that the properties of a galaxy also depend on its formation history and the environment, where it is embedded. 

Observational studies have also hinted at a connection between galaxy and group properties and the large-scale environments wherein they reside: Equal mass groups in superclusters have higher correlation function amplitudes \citep{2012MNRAS.426..708Y}; late-type brightest group galaxies have higher luminosities and stellar masses, redder colors, lower star formation activity, and longer star formation timescales when embedded in superstructures \citep{2015MNRAS.448.1483L}; the fraction of red galaxies is higher in high-density environments (superclusters) than in low-density environments \citep{2007A&A...464..815E,2014A&A...562A..87E}; at a fixed local density, galaxy colors change toward void walls \citep{2008MNRAS.390L...9C}; galaxies within groups with the same richness and mass are redder/older in high-density regions \citep{2012A&A...545A.104L,2013MNRAS.432.1367L}. These results support environmental effects but are not conclusive about the dependence between
large-scale environment and galaxy properties.

The stellar mass is a basic galaxy parameter that correlates strongly with almost all other galaxy properties \citep{2004MNRAS.353..713K} and also with the mass of its dark matter halo \citep{2010ApJ...710..903M}. Thus the stellar mass function (SMF) of galaxies is an observational quantity often used to constrain the models of galaxy formation and evolution. Recently, \citet{2015MNRAS.451.3249A} reported that  the SMF of galaxies also seems to vary in different large-scale structures. There is, however, no common consensus about its global shape and also its behavior in different large-scale environments is still an open question. This is largely associated with the nontrivial definition of large-scale environments and uncertainties in stellar mass estimations, which depend on the models and wavelength coverage used. The multifrequency study of galaxies from ultraviolet (UV) to infrared (IR) wavelengths allows us to constrain dust in galaxies and construct a more detailed and accurate picture of physical properties, including stellar masses, of a wide range of galaxy types. Recently, several large-scale sky surveys have been carried out at various wavelengths, e.g., in the far-ultraviolet (FUV) and near-ultraviolet (NUV) by the Galaxy Evolution Explorer (GALEX), in the optical by the Sloan Digital Sky Survey (SDSS), in the near-infrared (NIR) by the UKIRT Infrared Deep Sky Survey (UKIDSS), and in NIR and mid-infrared (MIR) by NASA’s Wide-field Infrared Survey Explorer (WISE). By combining the data from these different surveys, one can construct broadband spectral energy distribution (SED) of galaxies with appropriate galaxy models. The best state-of-the-art model available for this purpose is the model developed by \citet{2008MNRAS.388.1595D}. It is based on the stellar population synthesis model of \citet{2003MNRAS.344.1000B} and the two-component dust model of \citet{2000ApJ...539..718C}. By fitting the observed SEDs with models, we can estimate rest-frame luminosities and other physical properties related to stellar and dust contents in galaxies. This method was successfully applied to nearby star-forming galaxies from the Spitzer Infrared Nearby Galaxy Survey (SINGS), 3258 low-redshift galaxies from the SDSS Data Release 6 \citep{2010MNRAS.403.1894D}, 250~$\mu$m selected galaxies from Herschel-ATLAS Survey \citep{2012MNRAS.427..703S}, and the Andromeda galaxy \citep{2014A&A...567A..71V}. It was also tested with simulated galaxies from hydrodynamical simulations \citep{2015MNRAS.446.1512H}.

Physical mechanisms like strangulation \citep{1980ApJ...237..692L}, ram-pressure stripping \citep{1972ApJ...176....1G}, and harassment
\citep{1996Natur.379..613M}, which are believed to transform galaxy properties on group and cluster scales, mainly operate on satellite galaxies. On much larger scales, the number of satellite galaxies in groups are found to vary depending on whether they are present within or outside filaments \citep{2015ApJ...800..112G}. Any interplay between the inner group and large-scale environments may influence the efficiency of typical physical mechanisms occurring in groups and result in differences observed in properties of satellite galaxies in different large-scale environments. The interdependence of environment and central and satellite galaxies on cosmological scales still remains a less explored topic in galaxy evolution. We take a novel step toward understanding the large-scale environmental effect on central and satellite galaxy evolution using multifrequency observations and galaxy SED modeling. In this paper, we study the differences of the galaxy SMFs for different combinations of the large-scale structure, central and satellite galaxies, and galaxy morphology. For this purpose, we first expand the optical SDSS DR10 galaxy and group catalog \citep{2014A&A...566A...1T} to IR and UV wavelengths with data from different large-scale sky surveys. Using a publicly available MAGPHYS SED fitting code, we determine galaxy rest-frame luminosities and stellar masses in groups inferred from galaxy data at wavelength range 0.15~$\mu$m to 12.33~$\mu$m. Then, we divide the galaxy $r$-band luminosity density field into low- and high-density, large-scale environments and study SMFs of central and satellite galaxies in these two distinct environments. Our data analysis also gives the luminosity functions of galaxies for 15 wavebands from UV to IR. Plots of these are shown in appendix A. 

\section{Data}
\subsection{Ultraviolet, optical, and near-infrared data}
Photometric data from UV to NIR wavelengths come from the GAMA Data Release 2 \citep{2015MNRAS.452.2087L}. The UV photometric data come from the GalexCoGPhot catalog, which provides GALEX $NUV$ (0.152~$\mu$m) and $FUV$ (0.231~$\mu$m) curve-of-growth photometry at optical positions for all GAMA DR2 objects. Optical and NIR data come from SersicCatAll catalog, which provides photometry as a result of fitting a single-Sersic (1-component only) profile to every GAMA DR2 object in each of the bands $u$ (0.3562~$\mu$m), $g$ (0.4719~$\mu$m), $r$ (0.6185~$\mu$m), $i$ (0.75~$\mu$m), $z$ (0.8961~$\mu$m), $Y$ (1.0319~$\mu$m), $J$ (1.2511~$\mu$m), $H$ (1.6383~$\mu$m), and $K$ (2.2085~$\mu$m). This catalog is constructed using SIGMA v0.9-0 (Structural Investigation of Galaxies via Model Analysis) tool on SDSS and UKIDSS LAS imaging data \citep{2012MNRAS.421.1007K}. All magnitudes are expressed in AB mags.

\subsection{Mid-infared data}
The MIR data are taken from the latest data release from the WISE survey \citep{2010AJ....140.1868W}, which has mapped the whole sky with an angular resolution of 6.1$\arcsec$, 6.4$\arcsec$, 6.5$\arcsec$, and 12.0$\arcsec$ in four wavebands, $W1$, $W2$, $W3$, and $W4$ with effective wavelengths 3.4, 4.6, 12, and 22~$\mu$m, respectively. In the unconfused regions on the ecliptic, WISE has achieved 5$\sigma$ point source sensitivities better than 0.08, 0.11, 1 and 6 mJy  in the four bands and a sensitivity better than 100 times than that of IRAS in the 12~$\mu$m band.

\subsection{Optical group catalog}
The optical group catalog used in this work is constructed from SDSS DR10 galaxies using a friends-of-friends algorithm and a linking length that varies with redshift as explained in detail in \citep{2014A&A...566A...1T}. The catalog consists of both isolated (not a member of any group) and grouped galaxies in all of the 588193 galaxies and 82458 groups. The effective area of coverage of the catalog is around 7221 square degrees and extends up to redshift 0.2. The catalog is flux limited with magnitude limit 17.77 mag in the SDSS $r$ band. The redshifts of galaxies are taken from the SDSS spectroscopic sample, the Two-degree Field Galaxy Redshift Survey \citep[2DFRS;][]{2001MNRAS.328.1039C}, the Two Micron All Sky Survey \citep[2MASS;][]{2003AJ....125..525J,2006AJ....131.1163S} Redshift Survey \citep[2MRS;][]{2012ApJS..199...26H}, and the Third Reference Catalogue of Bright Galaxies \citep[RC3;][]{1994AJ....108.2128C}. The morphological classification of galaxies is based on the fraction of luminosity contributed by the de Vaucouleurs profile ($f_{deV}$), the exponential profile axis ratio ($q_{exp}$), $u-r$ and $g-r$ colors (see \citet{2011A&A...529A..53T} for details).

\subsection{Construction of the far-UV to mid-IR catalog}
The galaxies in the optical group catalog were first identified in GAMA, which resulted in 11056 galaxies and 1651 groups. In these SDSS groups, 380 galaxies  do not have GAMA counterparts and we use SDSS DR10 photometric information for these galaxies. Combining these galaxies, the matched catalog resulted in 11436 galaxies and 1651 groups. GAMA DR2 already provides cross identifications between UV, optical, and NIR data. To combine MIR data with other wavebands, we searched for all the WISE objects lying within an angular distance of 3$\arcsec$ from optical positions of galaxies in the matched catalog. This angular distance was chosen after many cross-matching trials so as to minimize the number of multiple associations and maximize the number of unique associations. Around 1.32$\%$ and 97.56$\%$ of the galaxies were found to have multiple and unique matches with WISE objects, respectively. All WISE objects that are flagged as contaminated by artifacts
in one or more WISE bands were deleted. This is about 3$\%$ of all matched WISE objects. Multiple associations were removed in the galaxy SED fitting process by selecting the match that fits better. We restricted our analysis in the redshift range 0.01--0.2. The final galaxy catalog consisted of 1635 groups and 11330 galaxies (about 2$\%$ of the SDSS DR10 galaxy catalog) containing photometric information in GALEX, SDSS, UKIDSS, and WISE wavelengths with the group membership criteria remaining the same as in the optical group catalog. The number of detected galaxies in different wavelengths are shown in column~8 of Table~\ref{sample selection}. The effective area of coverage of the catalog is about 144 square degrees. It includes all the three equatorial survey regions of GAMA I called G09, G12, and G15.
\subsection{Quantifying galaxy environments} 
The final step in our data preparation involved quantification of large-scale environments in the region of our catalog. Segregating different large-scale environments is not a trivial task, since filaments and superclusters are mostly unvirialized structures and their density is only slightly larger than the cosmic mean density \citep{2012A&A...539A..80L}. As is well-known, most of the galaxies are small and have a low luminosity. Hence, a simple number density analysis of galaxies, which counts all the galaxies with the same weight, cannot provide a precise description of the cosmic web. The luminosity density field of galaxies is a powerful and widely used technique to define the large-scale density distribution. In this method, the galaxy luminosities, corrected by observational biases and selection effects, are smoothed with an appropriate kernel, the width of which determines the characteristic spatial scale. This approach leads to the total three-dimensional luminosity density field of the survey, where different density levels represent different characteristic structures: clusters, filaments, voids, and superclusters. Following \citet{2012A&A...539A..80L}, we use the normalized optical galaxy luminosity density field with a smoothing scale of 8 $h^{-1}$ Mpc for characterizing the large-scale environment of our galaxy and group sample. We use the optical galaxy density value of 1.5 in units of cosmic mean luminosity density to divide our galaxy sample into low- and high-density, large-scale environments containing 31$\%$ and 69$\%$ of all galaxies in the sample, respectively. This value was also used in \citet{2009A&A...495...37T} to classify galaxies into voids and other large-scale environments. 
\section{Methods}
\subsection{Fitting spectral energy distributions}
The UV to IR part of the SED contains a lot of information about stars and dust in galaxies. However, to extract such information, models are necessary  to link physical properties of a galaxy with its observed SED. The best state-of-the-art model available that is consistent with observations and simulations, is that by \citet{2008MNRAS.388.1595D}. This model is based on the principle that the light emitted from stellar populations in a galaxy is partly absorbed by dust and re-emitted at longer wavelengths. It computes the spectral evolution of galaxies using the population synthesis model of \citet{2003MNRAS.344.1000B} , whereas dust attenuation is computed using the two-component model of \citet{2000ApJ...539..718C}. We use a publicly available MAGPHYS SED fitting tool to fit the observed galaxy SEDs to a library of model SEDs extending from UV to IR wavelengths with known physical parameters. The fitting produces the best-fit, rest-frame magnitudes and various other physical parameters related to galaxies and also provides the range of their likelihood values using a Bayesian approach.

There are a few concerns about the model used in this work: the effects of viewing angle, AGN contamination, and dust grain composition. The recent work by \citep{2015MNRAS.446.1512H}, in which they performed MAGPHYS SED fitting on simulated galaxies from hydrodynamical simulation, has shown that MAGPHYS recovers most of the physical parameters well regardless of the viewing angle. They also found that even when the contribution of AGNs to the UV--mm luminosity is around 25 percent, the fits are acceptable and the parameters recovered are accurate. The largest discrepancy appears when assuming a different composition of dust. Both Milky Way type dust and Large Magellanic Cloud type dust produce acceptable fits, but the discrepancy increases for Small Magellanic Cloud type dust composition. Our sample consists of bright spiral galaxies and hence the Milky Way type dust is a reasonable assumption.     

Carefully calibrated photometric data for different wavelengths and from different surveys are needed for reliable estimates of SEDs. We use profile-fitted fluxes that give a measure of the total stellar light from a galaxy at all wavelengths as input to the code. In WISE bands, for the sources with signal-to-noise (S/N) ratios of less than two, the given magnitudes are 2 $\sigma$ brightness upper limits in magnitude units. Such sources account for 1.1$\%$, 1.1$\%$, 28.1$\%,$ and 68.0$\%$ of all galaxies in the matched catalog in $W1$, $W2$, $W3,$ and $W4$ bands, respectively. In WISE, photometric measurements of both stars and galaxies are based on profile fits using the point spread function as the source model. Although this method minimizes the effects of source blending, this may lead to underestimation of true fluxes for extended sources. \citet{2015ApJS..219....8C} found that the flux underestimation depends on the effective radius ($R_e$) of the source and may be estimated as
\begin{equation}
\triangle m = 0.10 + 0.46 \log(R_e) + 0.47 \log(R_e)^2 + 0.08 \log(R_e)^3     
.\end{equation}
The correction is applied to galaxies with effective radii of greater than 0.5 arcseconds. Following \citet{2015ApJS..219....8C}, we use the effective radii obtained after dividing SDSS $r$-band effective radii by a factor of 1.5 to convert galaxy size from optical to infrared. These corrections are added to flux uncertainties in the respective WISE wavebands.  
     
\begin{figure*}
\centering
                    {\includegraphics[width=\hsize]{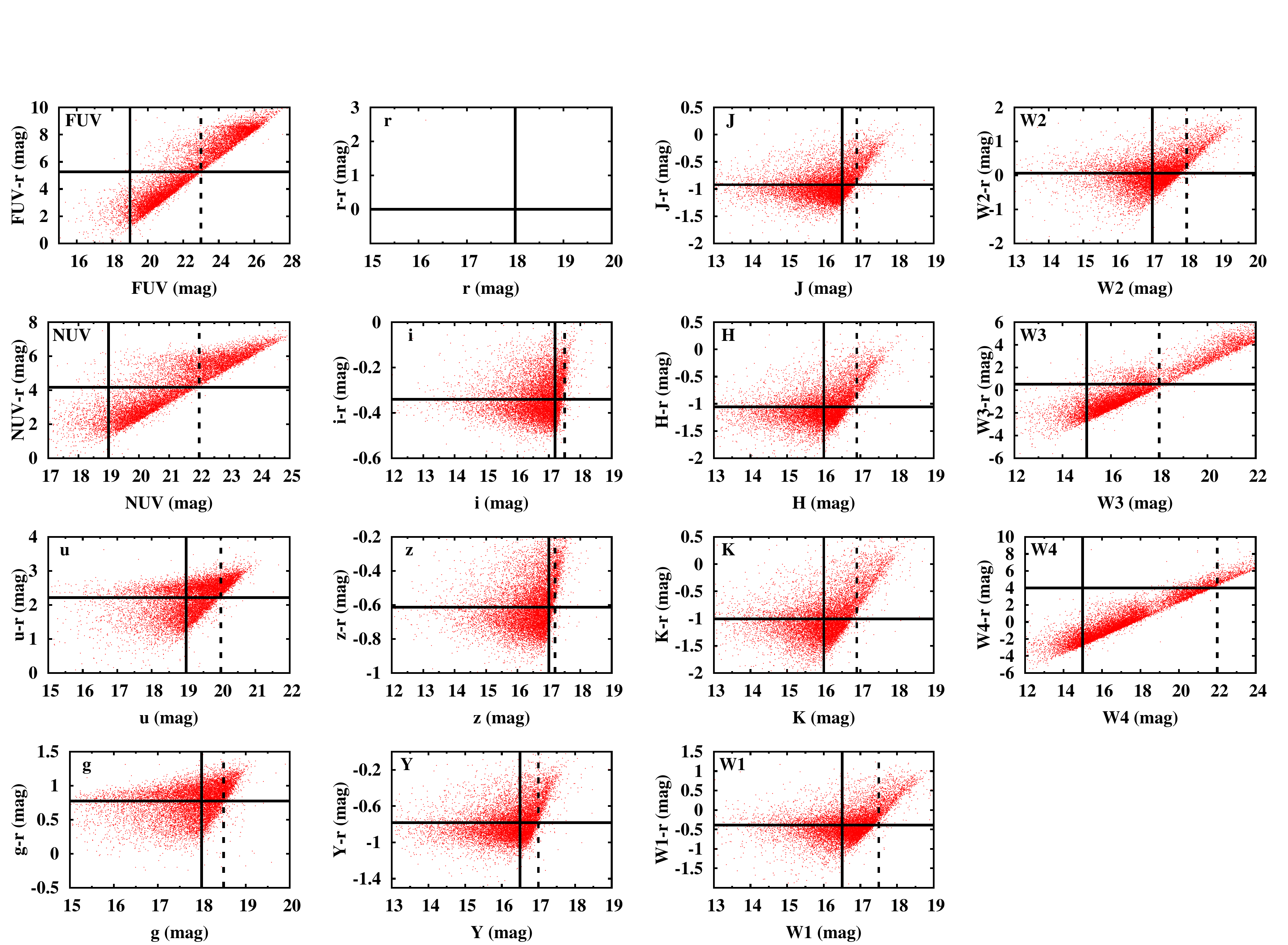}
      \caption{Color bias in the sample. The horizontal solid line shows the mean color in the indicated band. The dotted line shows the faint magnitude limit, while the vertical solid line shows the bright limit of the sample. The choice of the limits are based on visual inspection. 
              }
         \label{Color bias}}
   \end{figure*}
   
\subsection{Reducing the color bias in the sample}
Our sample is $r$-band selected ($m_r$ < 17.77 mag), so that the sample galaxies in the $r$ band do not represent a volume limited (VL) sample, but are affected by a certain amount of Malmquist-type bias. In the other wavelength bands, not originally used for selection, the biases generally differ \citep[cf. the "Gould effect";][]{1993ApJ...412L..55G}. Therefore, the form of the average SED of the observed sample differs from the average SED of a representative VL sample. This becomes visible as color biases in the observed sample, as shown in Fig.~\ref{Color bias}. As a result of this bias, galaxies with specific color are not detected in the sample as seen from the gap at the lower right part of the color magnitude plots. $FUV$, $NUV,$ and $W3$ bands are very strongly color biased. The sample of galaxies fainter than the bright limit (vertical solid line) have negligible color biases but this limit is not used as it reduces the sample size significantly for any statistical analysis. In order to reduce color bias in the sample, we adopt the method used by \citet{2012MNRAS.427.3244D} and calculate the mean color of the bands with respect to the $r$ band; we make a magnitude cut as shown in Fig.~\ref{Color bias} to remove all galaxies fainter than the faint limits (vertical dashed line) in each of the wavebands separately. Although this method does not completely remove the color bias, it is a good compromise between the bias and sample size. This procedure is applied after estimating rest-frame magnitudes from the SED fitting as was explained in the previous section. The number of galaxies in different wavebands after such color cuts is tabulated in column 9 of Table~\ref{sample selection}. 
 
\section{Fit results}
\begin{figure}[htbp]
   \centering
   \includegraphics[width=\hsize]{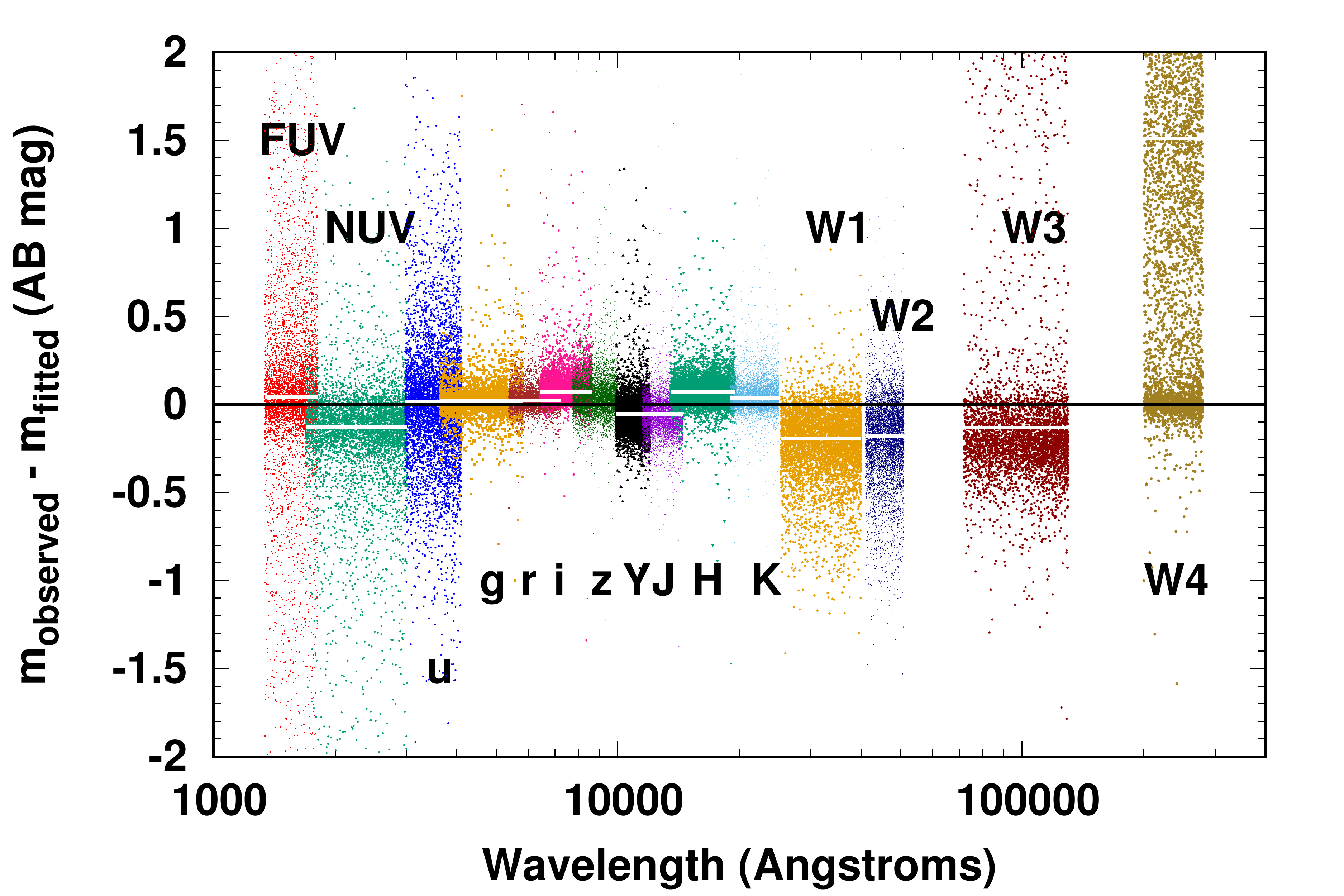}
      \caption{Difference between observed and MAGPHYS fitted magnitudes at different wavelengths. The white solid line shows the median values.
              }
         \label{magnitude difference}
   \end{figure}

Figure~\ref{magnitude difference} shows the difference between the observed and model fitted magnitudes at various wavebands used. The median difference is close to the zero line in all wavebands except $W4$ band. In the $W4$~band, the median difference is about +1.5 mag. The upper limit in source fluxes in WISE bands may contribute to the large offset seen in the $W4$ band. The real signal in these bands may be much weaker than these upper limits, which result in offset toward positive values. Also, $W4$ filter transmission curve has been revised, but MAGPHYS models are not yet updated in the $W4$ band. Similar offset in $W4$ band is also seen in \citet{2016MNRAS.455.3911D}.

\begin{figure}[htbp]
   \centering
   \includegraphics[width=\hsize]{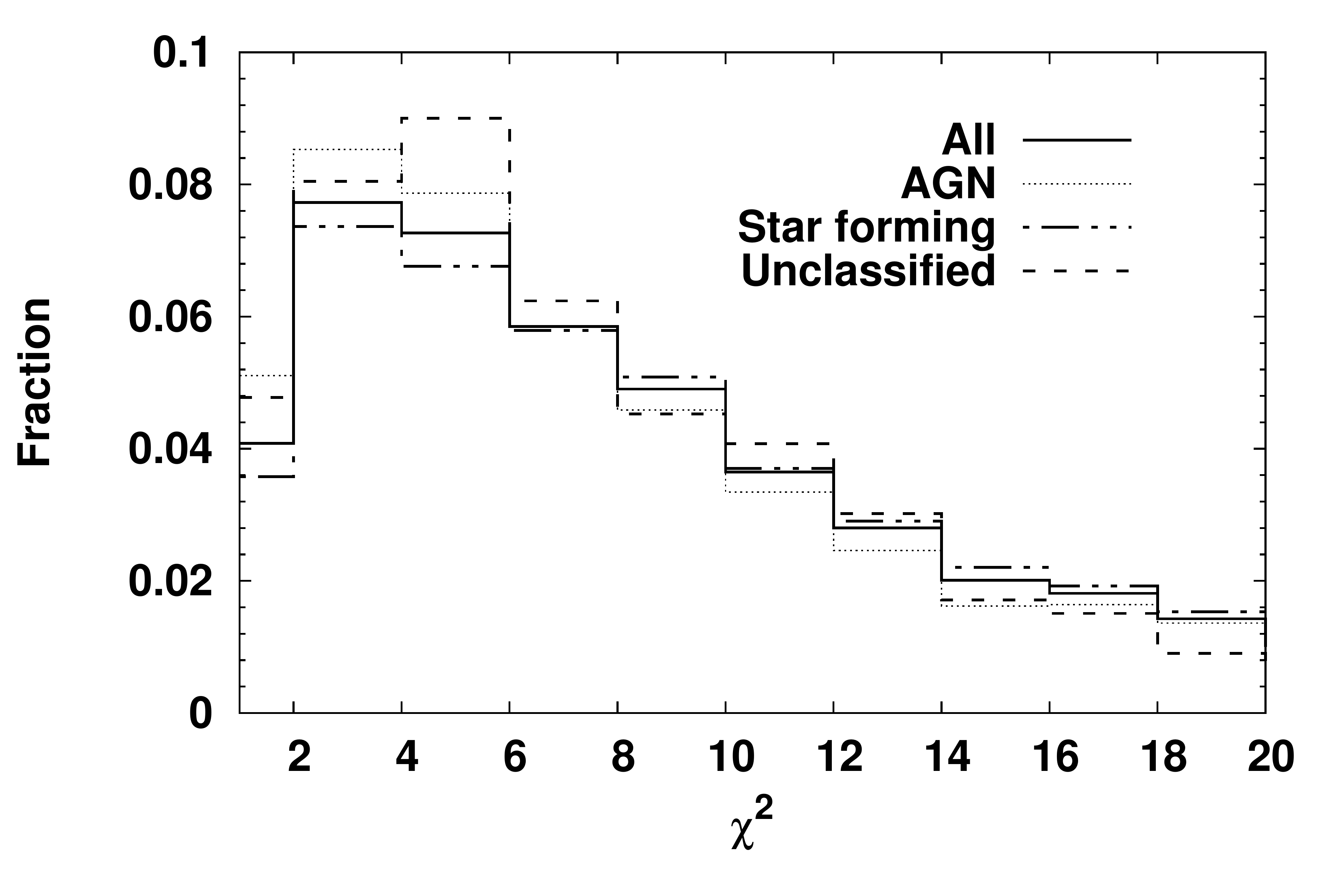}
      \caption{Chi square ($\chi^2$) goodness of fit values obtained from SED fitting. Different line types indicate GAMA classifications.
              }
         \label{chi square values}
   \end{figure}

The fitting also provides chi square ($\chi^2$) goodness of fit values. The $\chi^2$ values are reasonably small with few outliers. To test whether neglecting AGN heating in the model results in bad fits, we divided galaxies in the catalog into star-forming galaxies and AGNs based on classifications provided in GAMA, which uses the \citet{2001ApJ...556..121K} classification scheme. No significant dependence of the chi square ($\chi^2$) goodness of fit values on galaxy types was found and the distribution of $\chi^2$ values (Fig.~\ref{chi square values}) are very similar for both the star-forming galaxies and AGNs. Based on the analysis of the model SEDs of MAGPHYS, \citet{2012MNRAS.427..703S} derived a relation (equation B1) between the number of bands used during SED fitting and the number of degrees of freedom. Our 15 band photometry corresponds to 9 degrees of freedom, which means that above a chi square value 20 there is a probability of less than 1 percent that the observations are consistent with the models. We set this limit when making any statistical analysis with the sample. Within this limit, MAGPHYS was found to reproduce the galaxy parameters of simulated galaxies with good accuracy \citep{2015MNRAS.446.1512H}. As the WISE catalog contains both stars and galaxies, such a $\chi^2$ limit also reduces the stellar contamination present in the sample when matching galaxies with WISE sources. 


\begin{table*}
\caption{Sample selection.}             
\label{sample selection}      
\centering          
\begin{tabular}{c c c c c c c c c c c }     
\hline\hline       

Filter & Wavelength  & M$_\odot$& Bright limit\tablefootmark{a}  & Mean color & Faint limit\tablefootmark{b} & M$_X$ limit\tablefootmark{c} &   N(z)\tablefootmark{d} & N(color)\tablefootmark{e} & N($\chi^2$+color)\tablefootmark{f}  \\
($X$)     & ($\mu$m)    & (AB mag) &(AB mag)    & ($X-r$)     & (AB mag)           & (AB mag)       &          &          &                      \\
\hline                    
$FUV$         & 0.152      & 16.02      & 19.0          & 5.26       & 23.0 & $-11.5$,$-19.5$         & 7678  & 4931      & 3985 \\
$NUV$         & 0.231      & 10.18      & 19.0          & 4.17       & 22.0 & $-13.4$,$-20.0$     & 7948   & 5355     & 4260 \\
$u$           & 0.356      & 6.38       & 19.0          & 2.22       & 20.0 & $-13.4$,$-20.5$     & 10581  & 9008     & 7362 \\
$g$           & 0.472      & 5.15       & 18.0          & 0.77       & 18.5 & $-14.4$,$-22.0$         & 11328   & 10091   & 8317 \\
$r$           & 0.618      & 4.71       & 18.0          & 0.00       & 17.8 & $-14.4$,$-23.0$     & 11330  & 11043    & 9228 \\
$i$           & 0.750      & 4.56       & 17.2          & -0.34      & 17.5 & $-14.9$,$-23.0$         & 11316  & 10981    & 9179 \\
$z$           & 0.896      & 4.54       & 17.0          & -0.61      & 17.2 & $-15.4$,$-23.5$         & 11115  & 10465    & 8737 \\
$Y$           & 1.032      & 4.52       & 16.5          & -0.78      & 17.0 & $-15.9$,$-24.0$         & 10591  & 9739     & 8089 \\
$J$           & 1.251      & 4.57       & 16.5          & -0.92      & 16.9 & $-15.9$,$-24.0$         & 10569  & 9601     & 7985 \\
$H$           & 1.638      & 4.71       & 16.0          & -1.05      & 16.9 & $-16.4$,$-24.0$         & 10751  & 9950     & 8289 \\
$K$           & 2.208      & 5.19       & 16.0          & -1.01      & 16.9 & $-16.4$,$-24.0$     &  10717  & 9655    & 8055 \\
$W1$          & 3.379      & 5.94       & 16.5          & -0.38      & 17.5 & $-15.9$,$-23.5$         & 11230  & 10209    & 8572 \\
$W2$          & 4.629      & 6.61       & 17.0          & 0.06       & 17.8 & $-15.4$,$-23.0$     & 11230  & 9675     & 8098 \\
$W3$          & 12.33      & 8.40       & 15.0          & 0.53       & 18.0 & $-17.4$,$-24.5$         & 11230  & 8014     & 6706 \\
$W4$          & 22.00      & 9.10       & 15.0          & 3.92       & 18.0 & $-17.4$,$-24.5$         & 11230  & 10472    & 8728 \\

\hline                    
\end{tabular}
\tablefoot{
\tablefoottext{a}{Apparent magnitude limit where the color bias is negligible.}
\tablefoottext{b}{Adopted apparent magnitude limit used for calculating luminosity and stellar mass functions of galaxies.}
\tablefoottext{c}{Absolute magnitude range where the luminosity function is reliable.}
\tablefoottext{d}{Number of galaxies that have detections in respective bands.}
\tablefoottext{e}{Number of galaxies in different bands after color cuts explained in section 3.2.}
\tablefoottext{f}{Number of galaxies in different bands after color and $\chi^2$ cuts (see section 4 for details).}
}
\end{table*} 
\subsection{Physical parameters from SED fitting}
\begin{figure}
   \centering
   \includegraphics[width=\hsize]{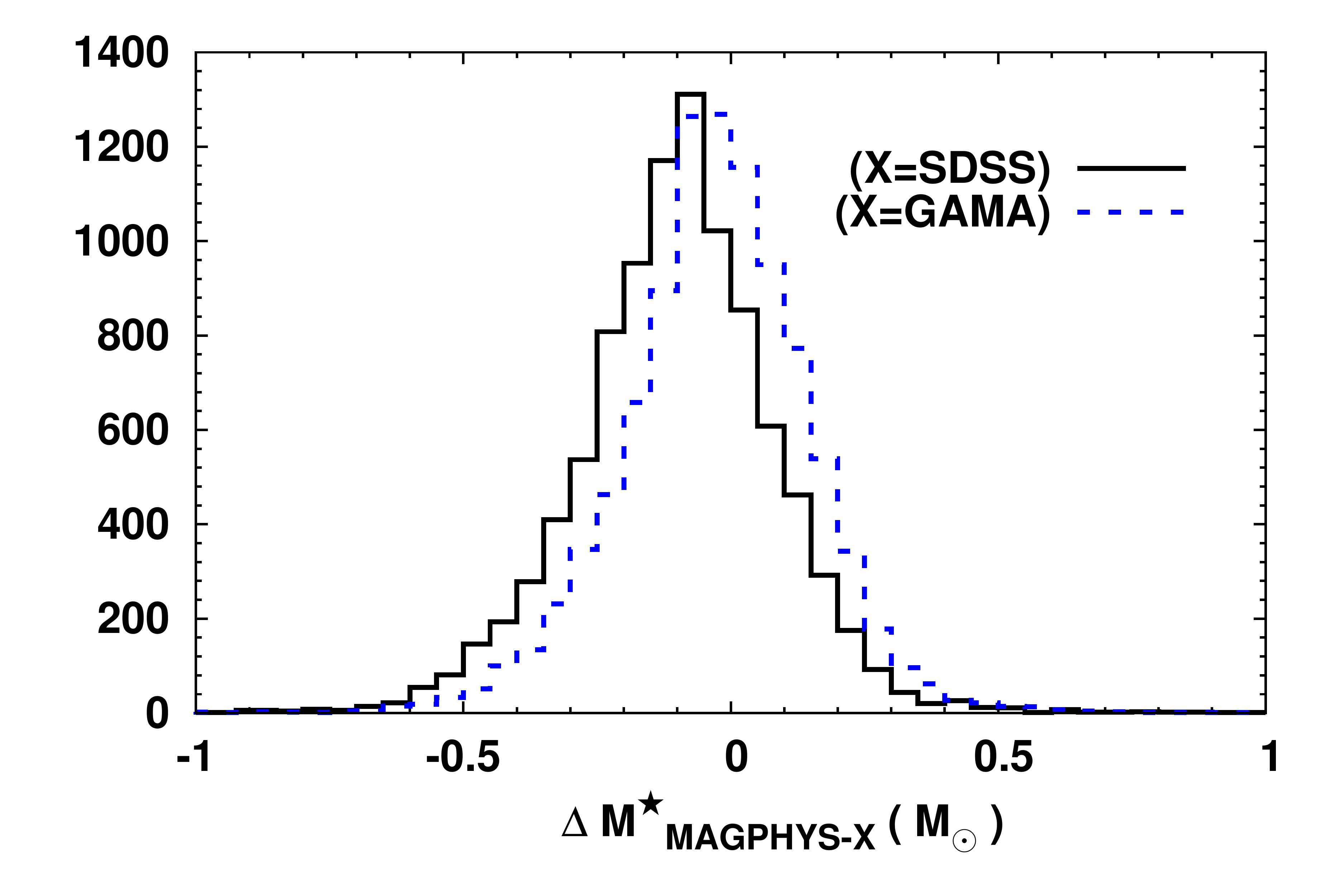}
      \caption{Histograms of differences between our stellar mass estimates (indexed as MAGPHYS) with SDSS (solid black) and GAMA (dashed blue) values. Masses are expressed in their log values. 
}
         \label{stellar mass}
   \end{figure}

\subsection{Stellar masses}
The fitting gives many physical parameters characterizing galaxies such as age, metallicity, stellar mass, and star formation rate. It also provides the likelihood distribution of stellar masses and we take median values as the best estimates. The stellar masses obtained agree very well with SDSS and GAMA \citep{2011MNRAS.418.1587T} values in general (Fig.~\ref{stellar mass}). The median and standard deviation of the differences between our and SDSS values are found to be $-0.098$ and $0.886,$ respectively. For GAMA, these values are found to be $-0.029$ and $0.477,$ respectively. For estimating stellar mass in galaxies, SDSS and GAMA use SED fitting to optical bands. The observed similarities show that using only optical band information can also provide good estimates for stellar masses in galaxies.

\subsection{Rest-frame absolute magnitudes}
The SED fitting also provides estimates of the expected rest-frame fluxes in different bands by simultaneously fitting the observed SEDs with models. It also provides flux estimates of galaxies in bands where detection has not been possible because of observational limitations. However, we only use rest-frame fluxes in bands which have detections and restrict our analysis to galaxies with $\chi^2 < 20$. Such a restriction increases the reliability of our results. We convert the rest-frame (k-corrected) fluxes in different bands obtained from SED fitting into absolute magnitudes in AB units for our analysis. The absolute magnitudes are not dust de-attenuated.  

\section{Statistical properties of the sample}
In order to check the reliability of the sample and its rest-frame luminosities obtained from SED fitting, we construct the galaxy
stellar mass and luminosity functions in different bands. We compare our galaxy SMF with the results from \citet{2008MNRAS.388..945B}         and
\citet{2014MNRAS.444.1647K}. We also compare the luminosity functions in different bands with SDSS \citep{2003ApJ...592..819B} and GAMA \citep{2012MNRAS.427.3244D} results. We use the Schechter fit parameters in the Table.~4 in \citet{2012MNRAS.427.3244D} for luminosity function comparisons.
\subsection{Galaxy luminosity function}
The galaxy LFs at all bands are estimated using the $1/V_\mathrm{max}$ statistical method \citep{1976ApJ...203..297S} along with some modifications made by \citet{2012MNRAS.427.3244D}. This method minimizes the effect of the color bias in the sample by adjusting the $1/V_\mathrm{max}$ volume according to each object's color. However, it is less effective at the low luminosity end, where a galaxy with a specific color may not be detected in our sample. At a waveband $X$ ($X$ = $FUV$, $NUV$, $u$, $g$, $i$, $z$, $Y$, $J$, $H$, $K$, $W1$, $W2$, $W3$ or $W4$), the maximum volume available for each galaxy in the sample is estimated using the magnitude limit as either $17.77 - (mag_r - mag_X)$ or the faint limit in that band, depending on which is brighter. The luminosity distribution is then obtained by
\begin{equation}
\mathrm{\phi (\mathit{M})} = \frac{c}{\eta} \mathrm{\sum_i \frac{I_{\mathit{(M,M+dM)}} (\mathit{M_i})}{\mathit{V_\mathrm{max,i}}}},
\end{equation}
where $c$ is the ratio of the number of galaxies before $\chi^2$ cut to the number of galaxies after $\chi^2$ cut. $V_\mathrm{max,i}$ is the maximum comoving volume over which the $i^{th}$ galaxy can be observed, I$_A(x)$ is the indicator function that selects the galaxies belonging to a particular absolute magnitude bin (taken here as 0.5 AB mag) and the sum runs over all galaxies in the sample. $\eta$ is the cosmic variance correction factor \citep[$\eta$ = 0.85,][]{2011MNRAS.413..971D}. For each waveband, the absolute magnitude limits fainter than which the method fails to correct for the color biases are calculated using the bright limits indicated in Table~\ref{sample selection}. The faint absolute magnitude limit values at different wavebands are shown in column 7 of Table~\ref{sample selection}. The absolute magnitude bins that contain fewer than ten galaxies toward the brightest part of the magnitude distribution do not contribute significantly to the overall luminosity density. The bright absolute magnitude limit values at different wavebands are shown in column 7 of Table~\ref{sample selection}. 

Figure~\ref{lumf} shows the obtained luminosity functions as well as a comparison with the GAMA and \citet{2003ApJ...592..819B} results for redshift 0.1. Within the selection boundaries, the optical galaxy LFs agree well with \citet{2003ApJ...592..819B} results. Compared to our estimates, the GAMA LFs are shifted toward the brighter end at optical and NIR bands except the K band. At FUV and NUV wavelengths, the GAMA luminosity functions have higher amplitudes than our results. These differences are mainly due to different magnitude limits, redshift ranges, photometry, and sampling. The GAMA sample is about 1.6 magnitude deeper than \citet{2003ApJ...592..819B} and our samples. The GAMA results use r-band matched aperture photometry, \citet{2003ApJ...592..819B} use Petrosian magnitudes, whereas we use Sérsic profile fitted magnitudes to construct the LFs at optical and NIR wavelengths. Sérsic magnitudes are brighter than Petrosian and aperture magnitudes for galaxies with higher Sérsic indices. Sérsic LFs were found to be shifted toward the bright relative to the LFs based on Petrosian and aperture magnitudes in the r band by \citet{2011MNRAS.412..765H}. This shows that different magnitude types used affect the derived LFs noticably.

\subsection{Galaxy stellar mass function}
\begin{figure}[t]
   \centering
   \includegraphics[width=\hsize]{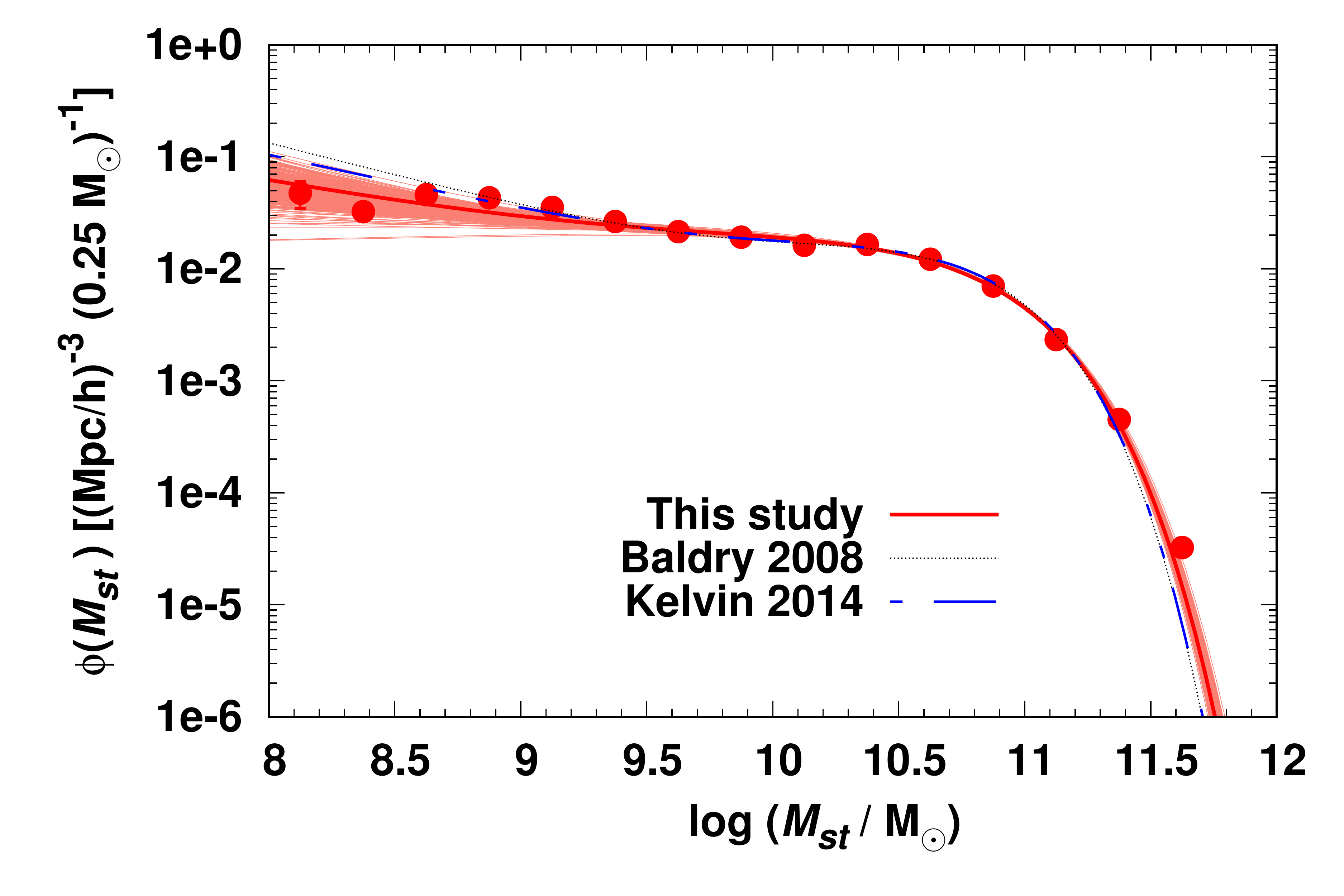}
      \caption{Stellar mass function of galaxies obtained from SED fitting and comparison with \citet{2008MNRAS.388..945B} (black dotted line)
 and \citet{2014MNRAS.444.1647K} (blue dashed line) results. The red solid line represents the best double Schechter function fit. The light red  lines, which form a shaded region, represent the individual double Schechter function fits to the galaxy stellar mass functions obtained from all bootstrap samples. The errors in density estimates at each stellar mass bins are calculated using bootstrap resampling technique.
}
         \label{stellar mass function: All galaxies}
   \end{figure}

 The galaxies in the sample are divided into logarithmic stellar mass bins to estimate the galaxy SMF. For each bin, the galaxy SMF is given by
\begin{equation}
\mathrm{\phi(\log \mathit{M_{st}}) d(\log \mathit{M_{st}})} = \frac {c}{\triangle \log \it M_{st}} \mathrm{\sum_\mathit{i} \frac{I_{\it(M_{st},M_{st}+dM_{st})} (\log \it M_{{st},i})}{\mathit{V_{max,i}}}} 
.\end{equation}\\
Here, $c$ is the ratio of the number of galaxies before $\chi^2$ cut to the number of galaxies after $\chi^2$ cut, and $V_{max,i}$ is the maximum comoving volume over which the $i^{th}$ galaxy can be observed and is obtained in the same way as explained in section 5.1. I$_A(x)$ is the indicator function that selects the galaxies belonging to a particular stellar mass bin (taken as $10^{0.5}$ M$_\odot$) and the sum runs over all galaxies in the sample. Recent studies have shown that the galaxy SMF has a distinctive bump around $10^{10.6}$ M$_{\odot}$ and can be best approximated by a double Schechter form with a combined knee \citep[see][]{2008MNRAS.388..945B,2010ApJ...721..193P,2012MNRAS.421..621B,2014MNRAS.444.1647K}. It can be expressed as
\begin{multline}
\mathrm{ \phi  (\log \mathit{M_{st}}) d(\log\mathit{M_{st}}) = \ln(10) exp(-10^{\log\mathit{(M_{st}/M_{st}^\star)}}) } \\
\mathrm{[\phi_1^\star   10^{\mathit{\log(M_{st}/M_{st}^\star)(\alpha_1 + 1)}} + \phi_2^\star 10^{\log\mathit{(M_{st}/M_{st}^\star)(\alpha_2 + 1)}}]  d(\log\mathit{M_{st})}}, 
\end{multline}
where $M_{st}^\star$ is the characteristic mass corresponding to the position of the distinctive "knee" in the mass function. The terms $\alpha_1$ and $\alpha_2$ are the slope parameters and $\phi_1^{\star}$ and $\phi_2^{\star}$ are normalization constants. The double Schechter function accurately models the bump observed in the galaxy SMF around $M_{st}^\star$, with one Schechter function dominating at stellar masses greater than $M_{st}^\star$  and the second one dominating at lower masses. The resulting galaxy SMF (red circular points) and the double Schechter fit including errors (red shaded region) are shown in the stellar mass range $10^{8.0}$--$10^{12.0}$ in Fig.~\ref{stellar mass function: All galaxies}. The curve fitting of the double Schechter function is performed using nonlinear least squares fitting technique with five free parameters. The fit limits and errors in number density of galaxies in each stellar mass bin are estimated using the bootstrap resampling technique. The fit parameters with
errors are shown in Table~\ref{double schechter stellar mass function}. The best-fit paramaters highly deviate from the results of \citet{2008MNRAS.388..945B} (B08 hereafter) and
\citet{2014MNRAS.444.1647K} (K14 hereafter). Considering the errors in fit parameters, $M_{st}^\star$ has higher values than B08 and K14. Both, $\phi_1^{\star}$ and $\alpha_1$ have lower values than B08 but lie within the errors when compared to K14 estimates. The remaining two parameters ($\phi_2^{\star}$ and $\alpha_2$) agree within the errors with both B08 and K14 values.

Unlike luminosity, which is a directly observable quantity, the stellar mass is commonly estimated from the multifrequency data by fitting the observed SEDs with a library of model SEDs with known physical properties. A number of different models exist regarding the initial mass function (IMF), star formation history (SFH), stellar populations, and dust that are used to build the model SEDs. Many studies have tried to quantify the effect of these assumptions on stellar mass estimates \citep{2009ApJ...701.1765M,2010MNRAS.407..830M,2010ApJ...709..644I,2012A&A...541A..85M,2013MNRAS.431.2209B,2013A&A...549A...4S}. These studies commonly agree that stellar masses can be reliably constrained but uncertainties up to 0.5 dex may exist depending on different models used. \citet{2009ApJ...701.1765M} quantified how the uncertainties associated with the assumptions made in the SED fitting method may affect the galaxy SMF. They find that the shape of the galaxy SMF at the low-mass end may change depending on the assumed metallicity and dust model. SDSS and GAMA  used only optical wavelengths to estimate stellar masses from the SED fitting. Here, we  attempt to include UV and MIR wavelengths. We do not find significant differences in our estimates although our values have small systematic offsets compared to SDSS and GAMA estimates. These differences can be attributed to a number of different factors such as inclusion of the UV and IR wavelengths, different assumptions in the SED fitting, and different photometry. These differences also affect the shape of the global galaxy SMF as seen in our results.

\begin{figure*}[htbp]
   \centering
   \includegraphics[width=\hsize]{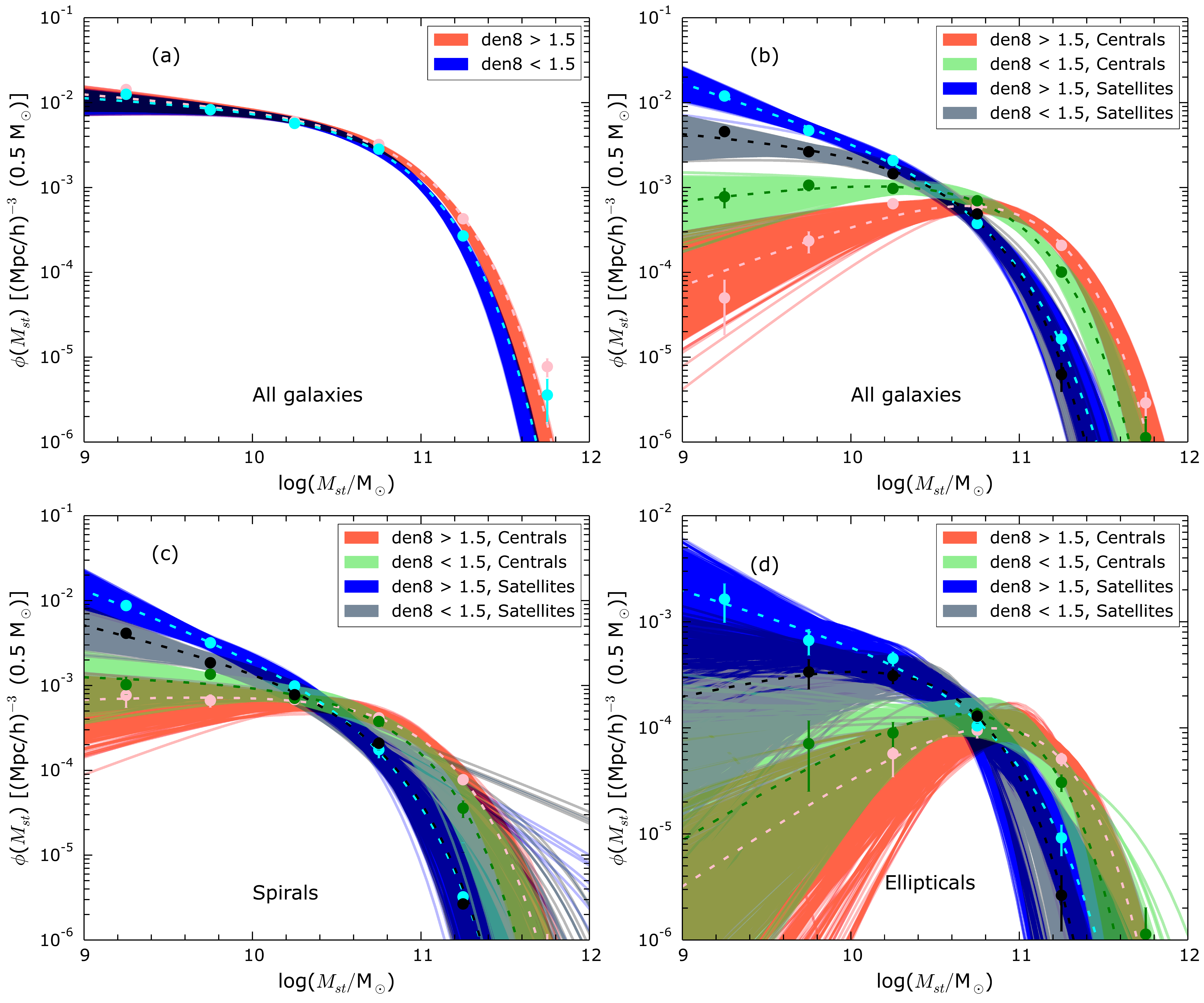}
      \caption{(a): Stellar mass function of galaxies in high- (light red points) and low- (sky blue points) density large scale environments. The light red and sky blue dashed
lines represent the best single Schechter function fits of galaxies in high- and low-density environments, respectively. The red and blue  solid lines, which form red (blue) shaded regions, show individual single Schechter function fits to the galaxy stellar mass functions obtained from all bootstrap samples of galaxies in high- and low-density environments, respectively. (b): Stellar mass function of central and satellite galaxies in high- and low-density, large-scale environments. The light red (sky blue) and dark green (black) dashed lines represent the best single Schechter function fits of central (satellite) galaxies in high- and low-density environments, respectively. The red (blue) and light green (gray)  solid lines which form red (blue) and light green (gray) shaded regions show individual single Schechter function fits to the  central (satellite) galaxy stellar mass functions obtained from all bootstrap samples in high- and low-density, large-scale environments, respectively. (c): Same as (b) but for galaxies with spiral morphologies. (d): Same as (b) but for galaxies with elliptical morphologies. In all of the plots, the points and error bars represent the mean densities and error in densities calculated using the bootstrap resampling technique.}
         \label{stellar mass functions: High and low density environments}
   \end{figure*}   
\begin{figure*}[htbp]
   \centering
   \includegraphics[width=\hsize]{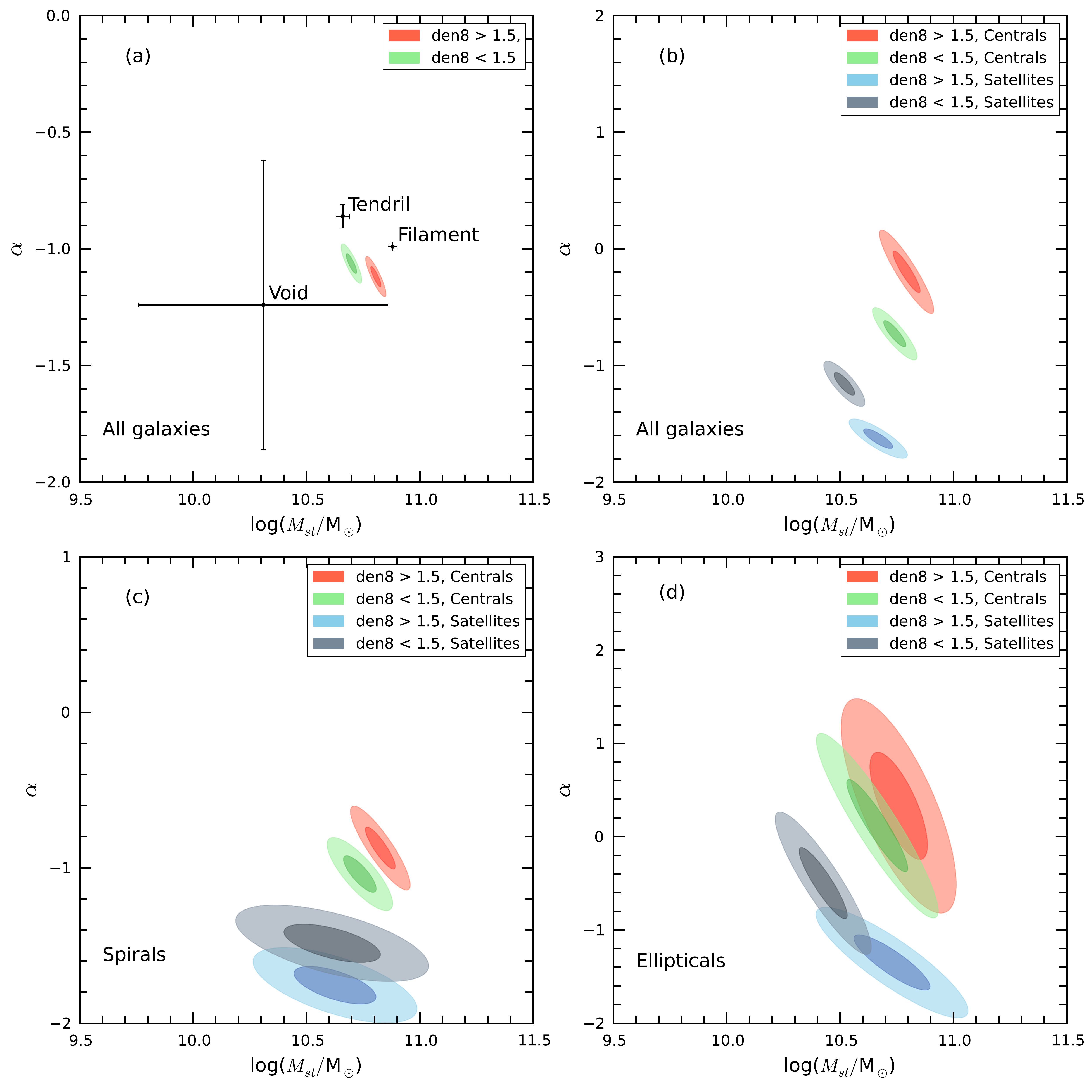}
      \caption{(a): Error ellipses at 95$\%$ confidence level for stellar mass function fit parameters ($M_{st}^\star$ and $\alpha$) of all galaxies in high (red ellipse) and low- (green ellipse) density, large-scale environments. The dark and light colored regions show 1$\sigma$ and 2$\sigma$ contours, respectively. The points with error bars show the results from \citet{2015MNRAS.451.3249A} in different large-scale structures, as indicated (b): Error ellipses at 95$\%$ confidence level for stellar mass function fit parameters ($M_{st}^\star$ and $\alpha$)
of central and satellite galaxies in high- and low-density, large-scale environments. The red (blue) and green (black) ellipses represent the error ellipses of central (satellite) galaxies in high- and low-density environments, respectively. The dark and light colored regions show 1$\sigma$ and 2$\sigma$ contours,
respectively. (c): Same as (b) but for galaxies with spiral morphologies. (d): Same as (b) but for galaxies with elliptical morphologies.
}
         \label{Error ellipses}
   \end{figure*}

\section{Galaxy stellar mass functions in different environments }

\subsection{All galaxies}\label{ssec:num1}
To study the large-scale environmental dependence on galaxy stellar mass, we divide our galaxy sample into low-density (Den8 < 1.5) and high-density (Den8 > 1.5) environments
according to their spatial location in the luminosity density field. We also construct the SMFs of galaxies in these two distinct environments. High-density environments have
a higher number density of galaxies than low-density environments. To take  this effect into account and to facilitate direct comparisons, we draw equal numbers of galaxies randomly
from each of these samples when constructing the SMFs. For each sample, we calculate the mean densities and error in densities in different stellar mass bins by combining all the SMFs obtained from 1000 different bootstrap subsamples and fit a single Schechter function to these values to get the best fit. For each sample, we also estimate all the single Schechter function fits to galaxy SMFs obtained from all the bootstrap subsamples. Figure~\ref{stellar mass functions: High and low density environments}(a) visualizes the results and Table~\ref{Single schechter stellar mass functions: All galaxies} shows the best-fit Schechter parameters with errors for galaxies in low- and high-density environments. Compared to low-density environments, the best Schechter function fit is shifted toward higher stellar masses for galaxies in high-density environments. In Fig.~\ref{Error ellipses}(a), we show the 1$\sigma$ (dark colored region) and 2$\sigma$ (light colored region) error contours of the corresponding Schechter fit parameters, characteristic stellar mass ($M_{st}^{\star}$) and low-mass slope ($\alpha$) at 95$\%$ confidence level. The SMFs of galaxies are different in high- and low-density environments. At 2$\sigma$, high-density environments have higher $M_{st}^{\star}$ than low-density environments, but $\alpha$ does not show any clear difference. For comparison, we also show the results (points with errorbars) from \citet{2015MNRAS.451.3249A} in three different environments, namely voids, tendrils, and filaments.These environments are defined using the filament finder algorithm based on a minimal spanning tree method \citep{2014MNRAS.438..177A}. As the definition of environment in our method is
completely different, our results can only be compared qualitatively with those in \citet{2015MNRAS.451.3249A}. Qualitatively, our results agree with \citet{2015MNRAS.451.3249A}; $M_{st}^{\star}$ increases toward high-density environments.

\subsection{All centrals and satellites}
We further divide each of our high- and low-density galaxy samples into centrals and satellites to study the environmental dependence on the stellar content of central and
satellite galaxies in groups in different large-scale environments. Central galaxies are those with the highest stellar mass in groups and the remaining members are classified as satellites. Our central galaxy samples also include isolated galaxies. We then construct the SMFs of centrals and satellites in both high- and low-density, large-scale environments (Fig.~\ref{stellar mass functions: High and low density environments}b). We use the same method as explained in section~6.1 to construct the galaxy SMFs. The corresponding single Schechter function fit parameters with errors are shown in Table~\ref{Single schechter stellar mass functions: All galaxies}. We also draw the 1$\sigma$ (dark colored region) and 2$\sigma$ (light colored region) error ellipses of $M_{st}^{\star}$ and $\alpha$ at 95$\%$ confidence level (Fig.~\ref{Error ellipses}b). For central galaxies, the error ellipses are well separated from each other, suggesting that the SMFs of central galaxies are different in high- and low-density environments. High-density environments have higher $\alpha$ value than low-density environments, but the $M_{st}^{\star}$ value shows no clear difference between different environments. Similar to central galaxies, the SMFs of satellite galaxies are different in high- and low-density environments. High-density environments have steeper $\alpha$ value than low-density environments. The $M_{st}^{\star}$ value is clearly higher in high-density environments at 1$\sigma$ only. The SMFs of central and satellite galaxies may have variations depending on the choice of the central galaxy definition and the group catalogs used (see appendix C for details).

\subsection{Centrals and satellites with fixed morphology}
 We divide central and satellite galaxy samples in both high- and low-density environments into spiral and elliptical populations to study the environmental dependence on the stellar content of central and satellite galaxies with fixed morphology in different large-scale environments. We further construct the SMFs of central and satellite galaxies with different morphologies across various large-scale environments using the method explained in section~6.1.
\subsubsection{Spiral morphology}
Figure~\ref{stellar mass functions: High and low density environments}(c) shows the SMFs of central and satellite galaxies with spiral morphologies in low- and high-density, large-scale environments as indicated. The single Schechter fit parameters are shown in Table~\ref{Single schechter stellar mass functions: Spiral galaxies}. The 2$\sigma$ error ellipses of $M_{st}^{\star}$ and $\alpha$ for central galaxies with spiral morphologies are  well separated (Fig.~\ref{Error ellipses}c). However, there is no clear difference in $M_{st}^{\star}$ and $\alpha$ values in high- and low-density environments within the error region considered. 

The 2$\sigma$ error ellipses of $M_{st}^{\star}$ and $\alpha$ for satellite galaxies with spiral morphologies are overlapping, but the 1$\sigma$ error ellipses are well separated (Fig.~\ref{Error ellipses}c). At 1$\sigma$, the $\alpha$ value is steeper in high-density environments than that in low-density environments. However, the $M_{st}^{\star}$ values for satellite galaxies are similar in both environments.
 
\subsubsection{Elliptical morphology}
Figure~\ref{stellar mass functions: High and low density environments}(d) shows the SMFs of central and satellite galaxies with elliptical morphologies in low- and high-density, large-scale environments as indicated. The single Schechter fit parameters are shown in Table~\ref{Single schechter stellar mass functions: Elliptical galaxies}. Both 1$\sigma$ and 2$\sigma$ error ellipses of $M_{st}^{\star}$ and $\alpha$ for central galaxies with elliptical morphologies in different environments are overlapping (Fig.~\ref{Error ellipses}d). There is no clear difference in $M_{st}^{\star}$ and $\alpha$ values in high- and low-density environments within the error region considered.

The 2$\sigma$ error ellipses of $M_{st}^{\star}$ and $\alpha$ for satellite galaxies with
elliptical morphologies are overlapping, but the 1$\sigma$ error ellipses are well separated (Fig.~\ref{Error ellipses}d). At 1$\sigma$, the $\alpha$ value is steeper in high-density environments than that in low-density environments. At 1$\sigma$, the $M_{st}^{\star}$ value for satellite galaxies is higher in high-density environments compared to that in low-density environments.

\section{Discussion}

\subsection{Galaxy stellar mass and halo relation}
According to the theory of galaxy formation, stars can form from cooling gas inside a virialized and gravitationally bound dark matter halo. The gas cooling and star formation rates of the galaxy depend mainly on the host halo mass \citep{1978MNRAS.183..341W}. In this scenario, the stellar mass of a galaxy can be expected to correlate strongly with the mass of its halo in which the galaxy formed. Several observational and simulation studies have tried to quantify the relation between the stellar mass of a galaxy and its halo mass. The stellar mass of a galaxy has been found to correlate with its halo mass with finite scatter. It increases rapidly with the halo mass when the halo mass is small, but slowly for large halo masses with a break in the relation around halo mass of 10$^{12}$ M$_{\odot}$ and galaxy stellar mass of 5 $\times$ 10$^{10}$ M$_{\odot}$ \citep{2010ApJ...710..903M}. The relation is very shallow at the high-mass end and the scatter in the relation is also higher. We note that our sample consists mainly of small groups and, thus, the correlation between the galaxy stellar mass and halo mass in our sample is expected to have less scatter. Some studies have also pointed out that the galaxy stellar and halo mass relation may differ between various morphologies and late-type galaxies reside in less massive halos than early types \citep[][and references therein]{2015ApJ...799..130R}. Using hydrodynamical simulations,
\citet{2015ApJ...812..104T} recently found a higher stellar to halo mass ratio in halos within the mass range 10$^{11}$--10$^{12.9}$ M$_{\odot}$ in large-scale denser environments. The dark matter halo mass function varies across different large-scale environments \citep{2007MNRAS.375..489H}. The low-mass end has the same slope in clusters, voids, filaments, and sheets, but the position of the high-mass cutoff depends on environment. The cutoff mass is the lowest in voids and gradually increases toward sheets, filaments, and clusters. In the halo mass range M < 5 $\times$ 10$^{12.9}h^{-1}$ M$_{\odot}$ , halos of a given mass tend to be older in high-density, large-scale environments and independent of environment at larger masses \citep{2007MNRAS.375..489H}. The present day halos in high-density environments are hence more massive and evolved than those in low-density environments as they have undergone many merger and accretion events during their evolution. The shift of the characteristic stellar mass toward higher stellar masses observed in our analysis of the SMFs of all galaxies in high-density, large-scale environments compared to the low-density environments may simply be the result of the mass and age segregation of halos in these environments.

\subsection{Stellar mass growth of central and satellite galaxies}
   \begin{figure}[t]
   \centering
   \includegraphics[width=\hsize]{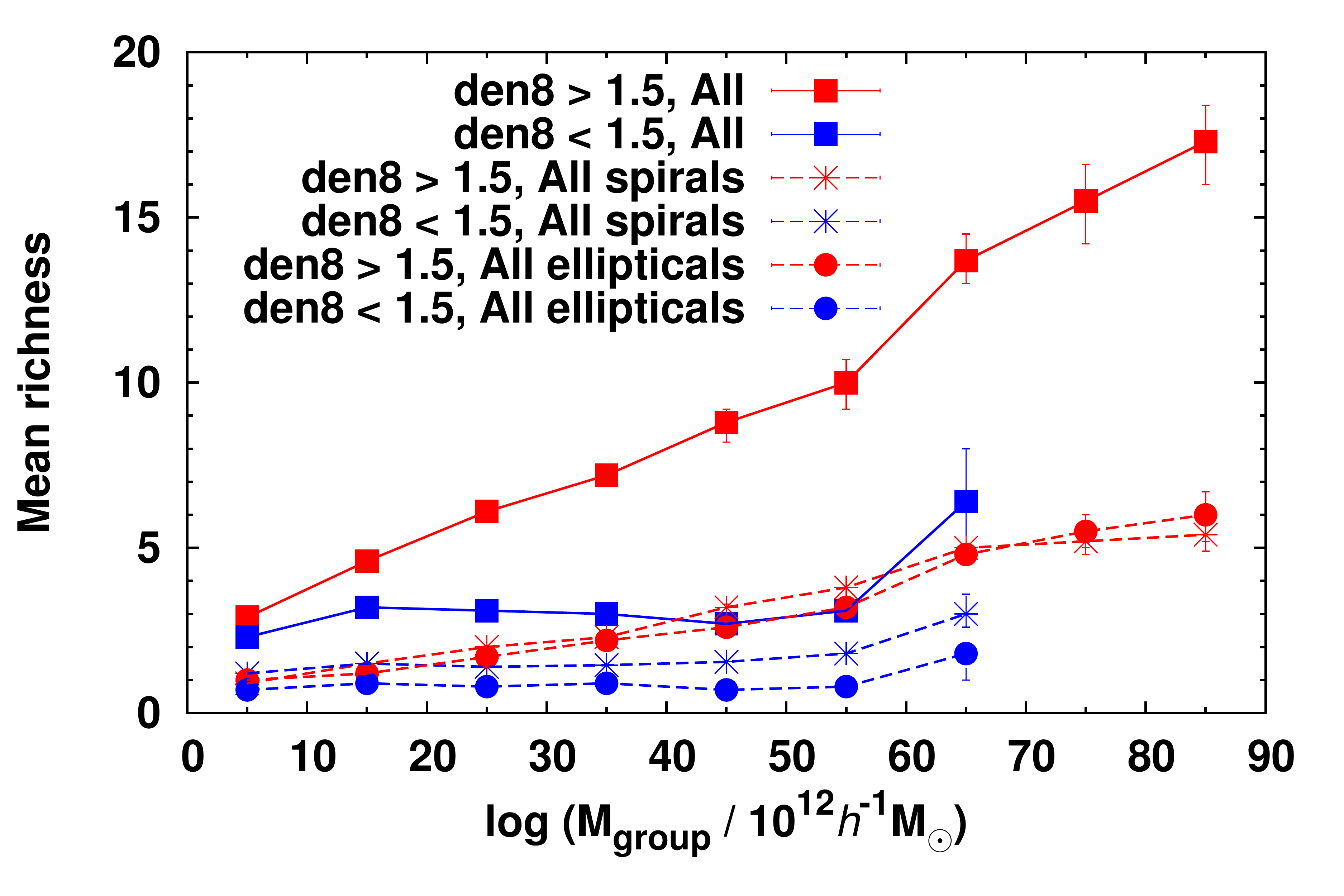}
      \caption{Mean richness of groups as a function of group dynamical mass. The red points and lines represent the relations in high-density scale environments, while blue points and lines represent low-density, large-scale environments. The square, circular, and asterisk points represent the results from all galaxy, spiral galaxy, and elliptical galaxy samples, respectively.}
         \label{mean richness vs group mass}
   \end{figure}
When a larger halo accretes a smaller halo with its central galaxy, the accreted halo turns into a subhalo and its galaxy becomes a satellite of the accreting halo. Stellar mass growth of galaxies then occurs mainly by star formation, accretion of smaller satellite galaxies, or major mergers. In low-mass halos, the stellar mass growth occurs mainly by star formation, while mergers play the dominant role in high-mass halos \citep{2012ApJ...746..145Z}. The galaxy merger rate in the semianalytical models is found to depend strongly on environment, where higher density regions have larger merger rates \citep{2012ApJ...754...26J}. According to our results, the characteristic stellar mass of the central galaxy SMF shows no clear differences in high- and low-density environments but the low-mass end slope is steeper in low-density environments. Such trends may be attributed to higher abundances of low mass and younger halos with low stellar mass central galaxies in low-density environments. In contrast to central galaxies, the satellite galaxies in high-density environments have higher characteristic stellar mass and the shift is only observed for satellite galaxies of elliptical morphologies and not for spiral satellites. Such differences may occur because galaxies in high-density environments reside in relatively more massive and evolved halos than low-density environments. The low-mass end in the SMFs of satellite galaxies have steeper slopes in high-density environments irrespective of their morphologies. This suggests that groups in high-density environments have higher abundances of satellite galaxies than those in low-density environments. We investigate further whether this is true at a fixed halo mass. We use the dynamical mass of a group as a proxy of its halo mass and study it as a function of group richness. It is clear from Fig.~\ref{mean richness vs group mass} that groups in high-density environments are richer in galaxies, i.e., show higher abundances of satellite galaxies in all dynamical mass bins than low-density environments. Similar trends are also seen in the richness of galaxies with spiral and elliptical morphologies. Recently, \citet{2015ApJ...800..112G} have also shown that central galaxies in filaments have more satellites than elsewhere. 

All these results support the scenario where groups with similar halos in high-density environments have high levels of substructures \citep{2007ApJ...666L...5E} and higher stellar mass content in those substructures. This suggests that galaxy formation is more efficient in subhalos located in high-density environments. A plausible scenario is that the high-density environments have more filaments, which supply cold gas to halos resulting in increased galaxy formation efficiency.

\section{Conclusion}
We have used GAMA DR2 data, WISE data, SDSS DR10 group catalog, and the MAGPHYS code to construct a multifrequency catalog of groups of galaxies with wavelengths ranging from UV to MIR. We studied the galaxy SMFs of different galaxy populations in groups in different large-scale environments. The main results are summarized below:

\begin{enumerate}
   \item
The global galaxy SMF obtained from multifrequency (0.152--22$\mu$m) galaxy SED fitting differs from results in the literature. It has a slightly higher characteristic stellar mass than results from \citet{2008MNRAS.388..945B} and \citet{2014MNRAS.444.1647K}.
\item
The SMFs of central galaxies in groups differ in high- and low-density environents. The low-mass end slope is steeper in high-density environments compared to low-density environments, but the characteristic stellar mass are similar in both environments.
\item
The SMFs of satellite galaxies in groups vary in different large-scale environments. High-density environments have a steeper low-mass slope and a higher characteristic
stellar mass than that in low-density environments.
\item
The SMFs of central galaxies with spiral morphologies are different in high- and low-density, large-scale environments. However, the differences in the Schechter fit parameters are not clear. The galaxy SMFs of central galaxies of elliptical morphologies are similar in high- and low-density, large-scale environments.
\item
The SMFs of satellite galaxies with spiral morphologies are different in high- and low-density, large-scale environments. In high-density environments the low-mass end slope is higher than that in low-density environments, but the characteristic stellar mass is similar in both environments. The SMFs of satellite galaxies with elliptical
morphologies are different in different large-scale environments. High-density environments have a higher low-mass end slope and a higher characteristic stellar mass
than that in low-density, large-scale environments.
   \end{enumerate}
   
Our first paper in the series of planned environmental studies of groups using multifrequency analysis of galaxies shows that the large-scale environment plays a significant
role in shaping the SMFs of different galaxy populations in groups. Groups in high-density environments have higher abundances of satellite galaxies, irrespective of the satellite galaxies morphology. For ellipticals, this trend is more obvious. Elliptical satellite galaxies are in general more massive in high-density environments. Stellar masses of spiral satellite galaxies do not show any dependence on large-scale environment. We also note that the SMFs of satellite and central galaxies may differ with group selection and central galaxy definition. The small sample size has restricted us to using a simple division of groups into void and nonvoid regions. However, the luminosity density method also allows us to separate high-density regions into finer structures such as filaments and superclusters. Using a much larger sample and wider wavelength coverage from the whole GAMA panchromatic data release \citep{2016MNRAS.455.3911D}, we plan to study stellar masses, luminosities, and star formation properties of groups located in these structures. This will allow us to constrain physical mechanisms by which large-scale environments may transform galaxy properties.

\begin{acknowledgements}
GAMA is a joint European-Australian project based around a spectroscopic campaign using the Anglo-Australian Telescope. The GAMA input catalog is based on data taken from the Sloan Digital Sky Survey and the UKIRT Infrared Deep Sky Survey. Complementary imaging of the GAMA regions is being obtained by a number of independent survey programs including GALEX MIS, VST KIDS, VISTA VIKING, WISE, Herschel-ATLAS, GMRT, and ASKAP providing UV to radio coverage. GAMA is funded by the STFC (UK), the ARC (Australia), the AAO, and the participating institutions. The GAMA website is http://www.gama-survey.org/.

This publication makes use of data products from the Wide-field Infrared Survey Explorer, which is a joint project of the University of California, Los Angeles, and the Jet Propulsion Laboratory/California Institute of Technology, funded by the National Aeronautics and Space Administration. AP acknowledges financial support from the
University of Turku Graduate School (UTUGS). HL acknowledges financial support from the Spanish Ministry of Economy and Competitiveness (MINECO) under the 2011 Severo Ochoa Program MINECO SEV- 2011-0187. ET and ME acknowledge the support from the ESF grants IUT26-2 and IUT40-2.      
\end{acknowledgements}

\bibliographystyle{aa} 
\bibliography{references.bib} 
\onecolumn
\begin{appendix}
\section{Luminosity function plots} 
\begin{figure*}[h]
   \centering
   \includegraphics[width=\hsize]{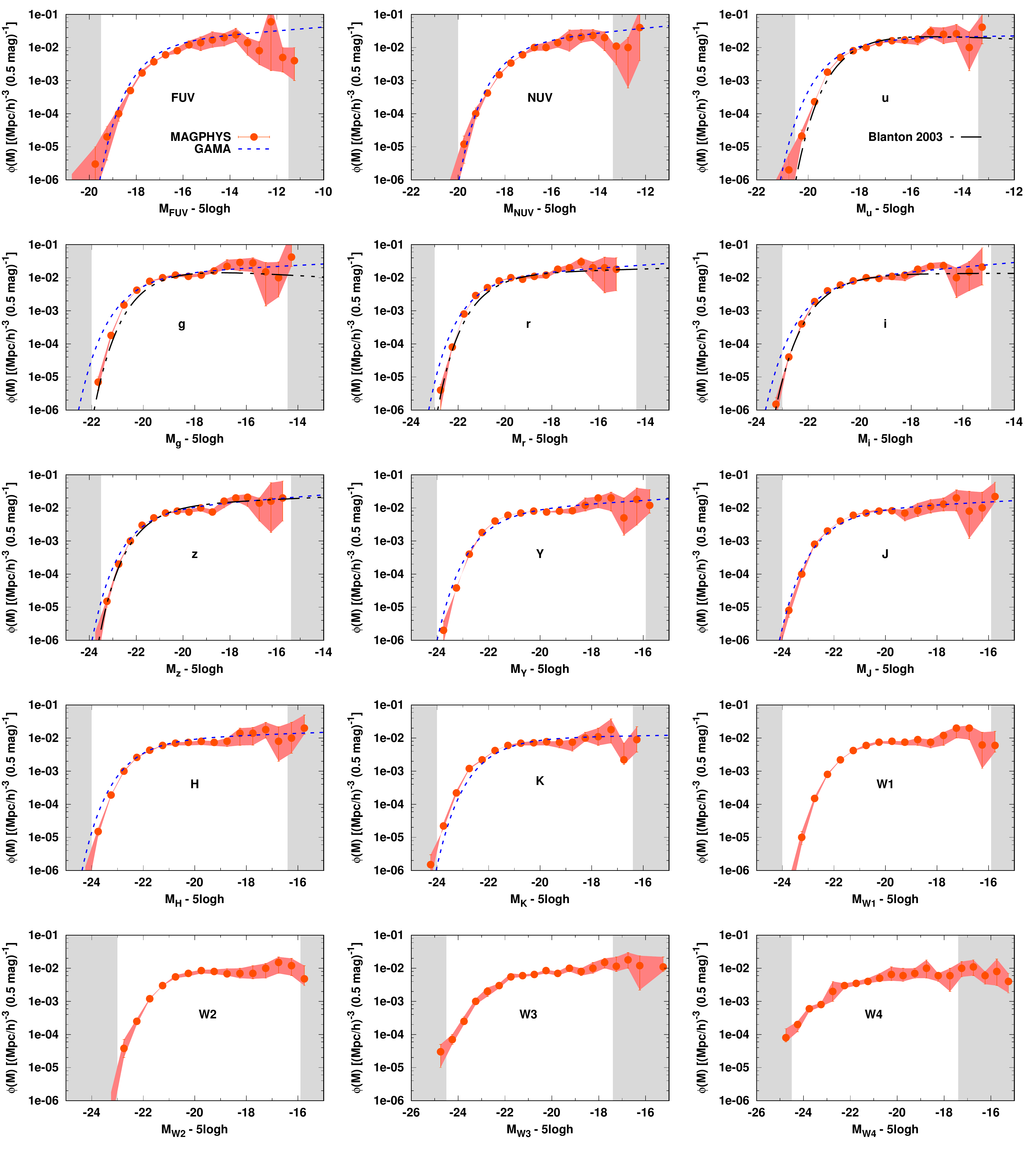}
      \caption{Luminosity functions of galaxies in $FUV$, $NUV$, $u$, $g$, $r$, $i$, $z$, $Y$, $J$, $H$, $K$, $W1$, $W2$, $W3,$ and $W4$ bands as indicated. The gray shaded region shows the selection boundaries explained in Section 4.2. The red points correspond to $1/V_\mathrm{max}$ corrected number densities in each absolute magnitude bin and the red shaded region shows the error limits. The blue dashed line shows GAMA luminosity functions and black dot-dashed line shows luminosity functions from \citet{2003ApJ...592..819B}. Both GAMA and \citet{2003ApJ...592..819B} estimate luminosity functions for galaxies up to redshift 0.1. Red points with error bars represent our estimates using $1/V_\mathrm{max}$ method in the redshift range 0.01--0.2.}
   \label{lumf}
   \end{figure*}
   
\section{Stellar mass function parameters}
\renewcommand{\arraystretch}{1.5}
\begin{table}[h]
\caption{Double Schechter stellar mass function fit parameters of all galaxies and comparisons with other results. }
\label{double schechter stellar mass function}
\centering
\begin{tabular}{lccccc}
\hline \hline
References& $\log {M_{st}^\star}$  & $\alpha_1$   & $\phi_1^{\star}$\tablefootmark{a} & $\alpha_2$ &  $\phi_2^{\star}$\tablefootmark{a}   \\
&(M$_\odot$)& & ($10^{-3}$X$^{-3}$) & & ($10^{-3}$X$^{-3}$) \\
\hline
This study& $10.73_{-0.01}^{+0.02}$&$-0.77_{-0.08}^{+0.13}$&$9.73_{-2.01}^{+1.06}$&$-1.43_{-0.09}^{+0.15}$& $1.66_{-0.92}^{+1.87}$\\
1& $10.64\pm{0.01}$&$-0.46\pm{0.05}$&$4.26\pm{0.09}$&$-1.58\pm{0.02}$& $0.58\pm{0.07}$ \\
2& $10.64\pm{0.07}$&$-0.43\pm{0.35}$&$4.18\pm{1.52}$&$-1.50\pm{0.22}$& $0.74\pm{1.13}$ \\
\hline
\end{tabular}
\tablefoot{
\tablefoottext{a}{X = Mpc in \citet{2008MNRAS.388..945B} and \citet{2014MNRAS.444.1647K}. X = Mpc$/h$ in this study}
}
\tablebib{
(1)~\citet{2008MNRAS.388..945B}; (2) \citet{2014MNRAS.444.1647K}.
}
\end{table}

\begin{table}[h]
\caption{Single Schechter stellar mass function parameters of galaxies in different samples.}
\label{Single schechter stellar mass functions: All galaxies}
\centering
\begin{tabular}{lccc}
\hline \hline
Sample & Subsample & $\log {M_{st}^\star}$  & $\alpha$  \\
&       &(M$_\odot$)    & \\
\hline
All galaxies& den8 > 1.5 & $10.85_{-0.07}^{-0.02}$&$-1.16_{-0.00}^{+0.08}$ \\
            & den8 < 1.5 & $10.75_{-0.08}^{-0.03}$&$-1.12_{-0.02}^{+0.10}$ \\
\hline            
Centrals    & den8 > 1.5 & $10.81_{-0.08}^{+0.04}$&$-0.25_{-0.12}^{+0.23}$ \\
            & den8 < 1.5 & $10.75_{-0.05}^{+0.04}$&$-0.76_{-0.08}^{+0.15}$ \\
\hline             
Satellites  & den8 > 1.5 & $10.67_{-0.07}^{+0.06}$&$-1.64_{-0.07}^{+0.10}$ \\
            & den8 < 1.5 & $10.52_{-0.05}^{+0.04}$&$-1.17_{-0.08}^{+0.11}$ \\                        
\hline
\end{tabular}
\end{table}

\begin{table}[h]
\caption{Single Schechter stellar mass function parameters of galaxies with spiral morphologies in different samples.}
\label{Single schechter stellar mass functions: Spiral galaxies}
\centering
\begin{tabular}{lccc}
\hline \hline
Sample & Subsample & $\log {M_{st}^\star}$  & $\alpha$  \\
&       &(M$_\odot$)    & \\
\hline
Centrals    & den8 > 1.5 & $10.85_{-0.09}^{+0.04}$&$-0.93_{-0.07}^{+0.19}$ \\
            & den8 < 1.5 & $10.74_{-0.12}^{+0.07}$&$-1.06_{-0.10}^{+0.14}$ \\
\hline             
Satellites  & den8 > 1.5 & $10.62_{-0.18}^{+0.19}$&$-1.77_{-0.10}^{+0.14}$ \\
            & den8 < 1.5 & $10.58_{-0.18}^{+0.24}$&$-1.48_{-0.13}^{+0.12}$ \\                        
\hline
\end{tabular}
\end{table}

\begin{table}[h]
\caption{Single Schechter stellar mass function parameters of galaxies with elliptical morphologies in different samples.}
\label{Single schechter stellar mass functions: Elliptical galaxies}
\centering
\begin{tabular}{lccc}
\hline \hline
Sample & Subsample & $\log {M_{st}^\star}$  & $\alpha$  \\
&       &(M$_\odot$)    & \\
\hline
Centrals    & den8 > 1.5 & $10.81_{-0.18}^{+0.07}$&$0.06_{-0.30}^{+0.84}$ \\
            & den8 < 1.5 & $10.68_{-0.15}^{+0.12}$&$-0.04_{-0.33}^{+0.65}$ \\
\hline             
Satellites  & den8 > 1.5 & $10.77_{-0.21}^{+0.12}$&$-1.47_{-0.17}^{+0.42}$ \\
            & den8 < 1.5 & $10.46_{-0.14}^{+0.07}$&$-0.63_{-0.25}^{+0.52}$ \\                        
\hline
\end{tabular}
\end{table}
\section{Central and satellite galaxy stellar mass functions}
\begin{figure}[htbp]
   \sidecaption
   \includegraphics[width=12cm]{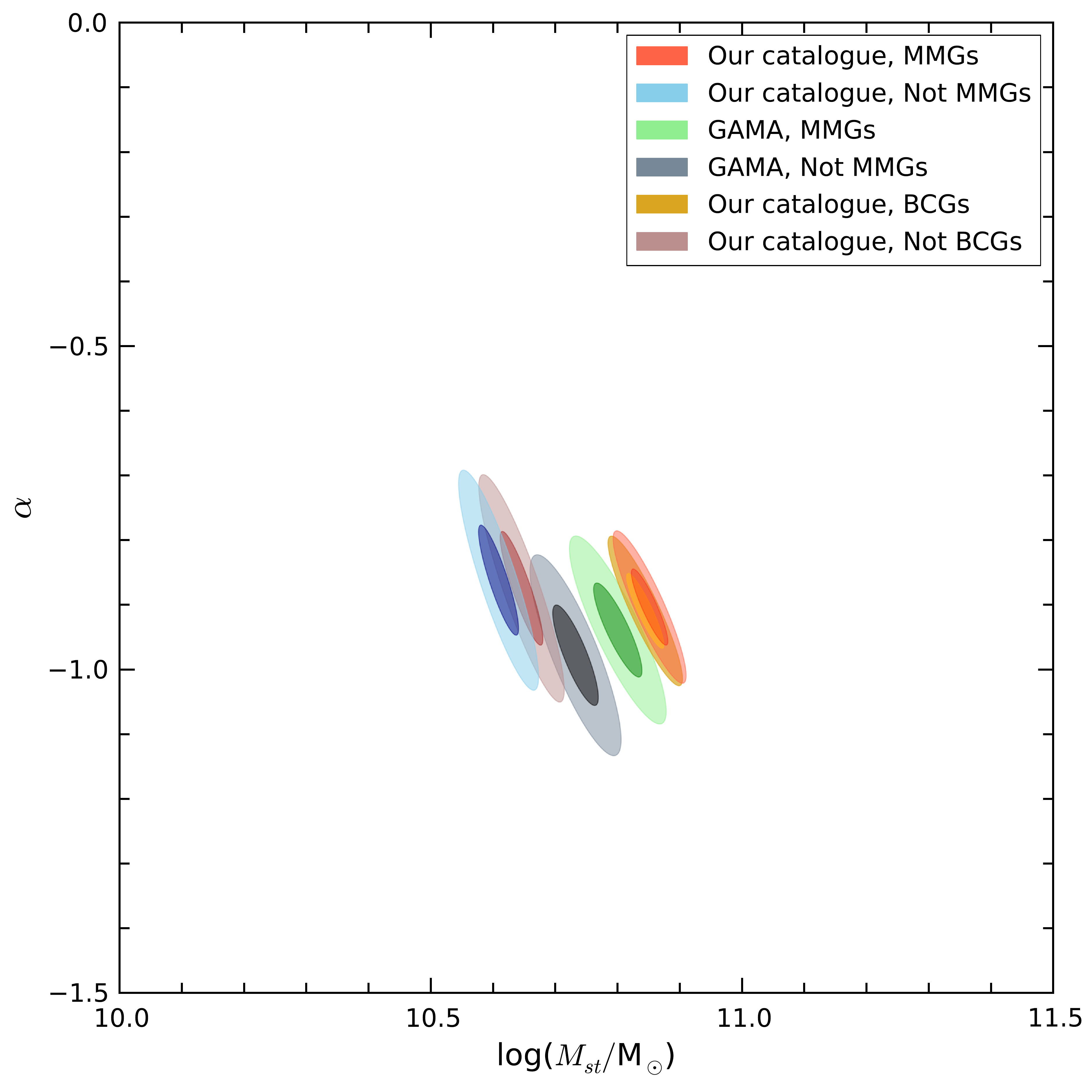}
      \caption{Error ellipses at 95$\%$ confidence level for stellar mass function fit parameters (${M_{st}^\star}$ and $\alpha$) of central and satellite galaxies for different catalogs and central galaxy definitions. MMGs represent the most massive galaxies in groups and BCGs represent the brightest galaxies in the groups. Red and blue (green and black) ellipses represent the error ellipses for the MMGs and not MMGs in our multifrequency (GAMA) group catalog. The golden and brown ellipses represent the error ellipses for the BCGs and not BCGs in our multifrequency group catalog, respectively.}
   \label{stellar mass function comparisons: GAMA and SDSS}
   \end{figure}

In this section, we perform a simple exercise to check how different definitions of central galaxies and use of different group catalogs may affect the SMFs of central and satellite galaxies. For this, we compare our results with that obtained from the GAMA group catalog \citep{2011MNRAS.416.2640R}. First, we find the galaxies in the whole GAMA catalog with the highest stellar mass in their groups and define these  as central galaxies and the remaining members as satellite galaxies. Then, we cross-match these galaxies with our multifrequency catalog, in which the central galaxies also have the same definition. Then, we extract two samples, one containing galaxies that are defined as centrals in the GAMA group catalog and other consisting of galaxies that are defined as centrals in our group catalog. Next, we make 1000 different random samplings from each of these samples and estimate the corresponding SMFs. In both cases, we use the stellar mass estimates from our multifrequency catalog. The red and green ellipses shown in Fig.~\ref{stellar mass function comparisons: GAMA and SDSS} represent the error ellipses of Schechter fit parameters, $M_{st}^\star$ and $\alpha$, for central galaxies in our and GAMA catalogs, respectively. The error ellipses are well separated from each other. However, the low-mass end slopes are similar but $M_{st}^\star$ seem to have lower values in GAMA centrals. Similarly, we also obtain the error ellipses of the Schechter fit parameters, $M_{st}^\star$ and $\alpha$ for satellite galaxies in GAMA and our catalog separately. The 2$\sigma$ error ellipses are overlapping but the 1$\sigma$ error ellipses are well separated. The low-mass end slopes are similar, but the GAMA satellites  clearly have higher $M_{st}^\star$.

In order to see if the definition of central galaxy based on luminosity affects the SMFs of the satellite and central galaxies, we selected the brightest galaxy in our sample as the central galaxy and remaining members as satellites. The golden and brown ellipses in Fig.~\ref{stellar mass function comparisons: GAMA and SDSS} show the corresponding error ellipses of the Schechter fit parameters, $M_{st}^\star$ and $\alpha$ for central and satellite galaxies, respectively. The error ellipse for the brightest central galaxies is almost the same as that for the most massive central galaxies. The error ellipse for the satellite galaxies of the brightest central galaxies overlaps with that for the satellite galaxies of the most massive central galaxies. However, $M_{st}^\star$ and $\alpha$ have similar values. This shows that the SMFs of central and satellite galaxies may differ based on the definition of the central galaxy and the group catalog used.

   \end{appendix}

\onecolumn

\end{document}